\documentclass[pdflatex,sn-mathphys-num]{sn-jnl}

\usepackage{subcaption}
\usepackage{graphicx}%
\usepackage{multirow}%
\usepackage{amsmath,amssymb,amsfonts}%
\usepackage{amsthm}%
\usepackage{mathrsfs}%
\usepackage[title]{appendix}%
\usepackage{xcolor}%
\usepackage{textcomp}%
\usepackage{manyfoot}%
\usepackage{booktabs}%
\usepackage{algorithm}%
\usepackage{algorithmicx}%
\usepackage{algpseudocode}%
\usepackage{listings}%
\usepackage{lineno}

\theoremstyle{thmstyleone}%
\theoremstyle{thmstyletwo}%

\theoremstyle{thmstylethree}%

\begin{document}

\title[Article Title]{A method for enhanced gamma–proton discrimination with imaging atmospheric Cherenkov telescopes}

\author[1]{\fnm{Jianling} \sur{Liu}}\email{liujl@my.swjtu.edu.cn}

\author*[1]{\fnm{Hu} \sur{Liu}}\email{huliu@swjtu.edu.cn}

\affil*[1]{\orgdiv{School of Physical Science and Technology}, 
\orgname{Southwest Jiaotong University}, 
\orgaddress{\city{Chengdu}, \postcode{610031},\country{China}}}


\abstract{
\textbf{Purpose:} Very-high-energy gamma-ray astronomy is an important window to study the extreme astrophysical processes in the universe, and how to effectively distinguish between gamma photons and background signals (mainly protons) is a key technical challenge to realize very-high-energy gamma-ray detection. Spaceborne calorimeters record secondary particle spatial distributions, achieving a background rejection power of $10^{4} \sim 10^{5}$ near 1 TeV. Ground-based imaging Cherenkov telescopes measure Cherenkov photon angular distributions, yet they only reach a rejection factor of $\sim$10 in this energy range, leaving substantial room for optimization. \\\indent
\textbf{Methods:} We use CORSIKA and sim\_telarray to simulate H.E.S.S. detector responses. Photon emission positions are reconstructed from Cherenkov shower images. Inspired by analysis techniques for space-borne calorimeter, we propose new observables derived from the transverse distribution of Cherenkov photon emission positions to separate gamma-ray signals from proton backgrounds. We employ a likelihood-ratio method to combine single-telescope variables across multiple telescopes and compare its performance with conventional Hillas-based analysis.\\\indent
\textbf{Results:} The results show that observables built from the transverse distribution of Cherenkov photon emission positions are less sensitive to Cherenkov image size and outperform the Hillas \textit{width} parameter for single-telescope gamma–proton discrimination. Furthermore, the likelihood ratio combining \textit{width} yields substantially better performance than the conventional mean reduced scaled width: the improvement reaches roughly one order of magnitude near \(1~\mathrm{TeV}\), and this enhancement remains above 50\% for energies above \(10~\mathrm{TeV}\).
}




\maketitle

\section{Introduction}\label{sec1}
\label{sec1}
Very-high-energy gamma-ray astronomy offers a vital probe of extreme cosmic astrophysical processes, enabling unique investigations of high-energy sources including active galactic nuclei, pulsar wind nebulae, and gamma-ray bursts~\cite{gamma-ray_astronomy,gamma-ray_future}. A critical detection challenge arises from cosmic-ray composition:over 99\% of cosmic rays arriving at Earth are charged nuclei (roughly 90\% protons, 9\% helium, and the remainder heavier species)~\cite{cosmic-ray_database}, which dominate the background for gamma-ray observations. Efficient gamma–proton separation is therefore an essential technical requirement for successful very-high-energy gamma-ray detection.\\\indent
Two primary gamma–proton separation techniques are available: space-based direct detection and ground-based indirect detection. Space observatories deploy 3D electromagnetic calorimeters for direct gamma measurement and gamma/proton discrimination. They record full 3D secondary particle showers from gamma rays or protons, exploiting stark spatial differences to reach high gamma(electron)/proton rejection power of $10^4 \sim 10^5$(gamma selection efficiency divided by proton efficiency)~\cite{ATIC-ep,DAMPE-ep,AMS-electron}. A representative example is China’s DAMPE (“Wukong”) satellite, whose electromagnetic calorimeter comprises 14 alternating horizontal layers of bismuth germanate (BGO) crystals with 32 total radiation lengths. This 3D geometry precisely captures shower longitudinal evolution and transverse broadening. Its electron–proton discrimination relies on two core observables~\cite{ATIC-ep,DAMPE-ep}:
(1)Longitudinal shower leakage: fraction of energy deposited in the final calorimeter layer relative to total shower energy;
(2)Transverse shower spread: average shower width measured across calorimeter layers.
These observables enable efficient electron–proton separation from tens of GeV up to several TeV, extending the measurable electron energy spectrum to multi-TeV energies.\\\indent
Ground facilities mainly employ the Imaging Atmospheric Cherenkov Technique (IACT) for indirect very-high-energy gamma-ray observation~\cite{Hess, Magic, Veritas}. The atmosphere acts as a natural calorimeter, and the imaging telescope records angular distributions of Cherenkov photons, emitted by electron–positron secondaries from extensive air showers (EAS). These photon images encode secondary shower spatial features, analogous to spaceborne calorimeter measurements. Standard Hillas parameters characterise each telescope’s Cherenkov image for gamma–proton separation, with the \textit{width} parameter being particularly important~\cite{Hillas_pars}. To improve discrimination, the \textit{width} from each telescope is combined into a new observable, the mean reduced scaled width (MRSW)~\cite{hess_crab}: it normalises \textit{width} to uniform mean and RMS across all $R_p$ and \textit{size} (total Cherenkov photons recorded in a single telescope image) bins before averaging over triggered telescopes. Nevertheless, modern IACTs with several telescopes only yield a rejection factor of $\sim$10 near 1 TeV~\cite{hess-ep,veritas-ep}, far inferior to space-based electromagnetic calorimeters.\\\indent
In this work, we analyze the link between Cherenkov shower images and the spatial distribution of photon emission positions, then propose new observables analogous to those developed for DAMPE. A log-likelihood ratio technique integrates multi-telescope information, and we benchmark these approaches against conventional Hillas-based discrimination methods.\\\indent
This paper is structured as follows: Section \ref{sec2} outlines the simulation dataset, Section \ref{sec3} presents the methodology, Section \ref{sec4} evaluates event discrimination performance, Section \ref{sec_disc} discusses the discrimination results, and Section \ref{sec5} provides conclusions. \\\indent

\section{Simulation}
\label{sec2}
Cosmic-ray air showers are simulated with CORSIKA v7.7410~\cite{Corsika}, with Cherenkov photons generated using the IACT module. The observation height is set to 1800 m above sea level (the site of High Energy Stereoscopic System, denoted as H.E.S.S.), using the U.S. standard atmospheric model. \par
Gamma-ray and proton primaries are simulated over 1–10 TeV with uniform $\log_{10}E$ sampling. In this energy range, we generate $1.5 \times 10^6$ gamma events and $2 \times 10^8$ proton events. Two discrete energy bins at $\log_{10}(E/\text{GeV}) = 4.25$ and $4.5$ are also simulated, each containing $3.5 \times 10^5$ gamma and $3.5 \times 10^6$ proton events. \par
\begin{figure}[H]
\centering
\includegraphics[width=0.7\linewidth]{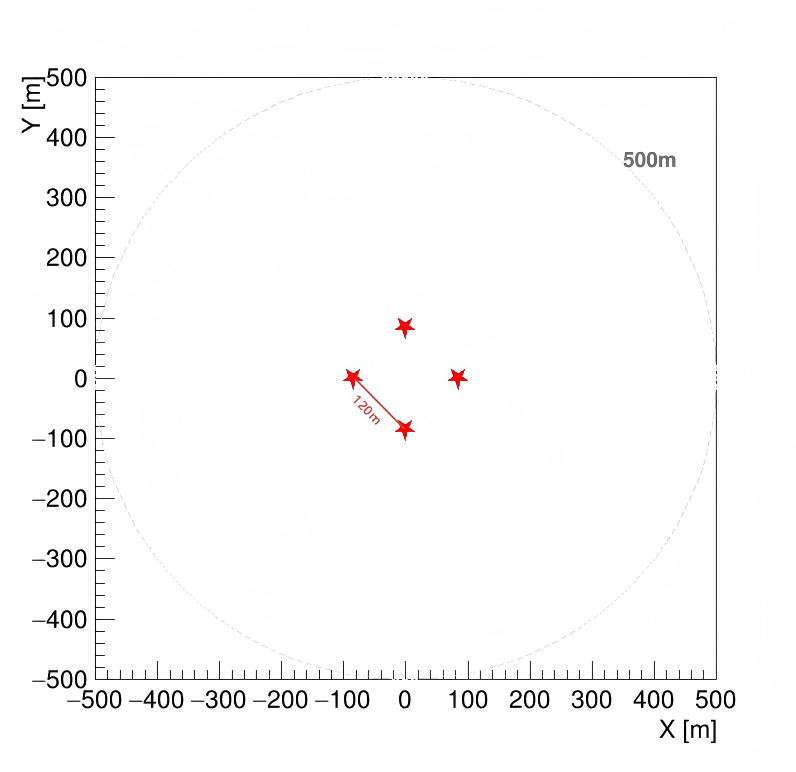}
\caption{Layout of the H.E.S.S. array, each star represents one telescope. Shower cores are randomly sampled within the dotted 500 m radius circle.}
\label{fig:enter-label}
\end{figure}

The zenith angle (hereafter $\theta$) is fixed to 20$^{o}$, yielding an atmospheric depth of $887~\text{g/cm}^2$ at the target altitude. The azimuth angle is set to 0$^{o}$. CORSIKA shower cores are randomly sampled within a 500 m radius circle, as shown in Fig. \ref{fig:enter-label}. The EPOS-LHC hadronic interaction model and EGS4 electromagnetic model are used for simulations. \par
We adopt the H.E.S.S. Phase-I telescope array for detector simulations. As shown in Fig.~\ref{fig:enter-label}, four telescopes sit at the vertices of a 120 m-sided square. Each telescope features a 12 m dish with 107 $m^{2}$ mirror area and a camera of 960 photomultiplier tubes; each pixel spans 0.16$^{o}$, yielding a 5$^{o}$ field of view. The response of the telescope array was simulated using the well-tested simtelarray package~\cite{simtelarray}. This simulation involves optical ray-tracing of the photons, their detection by photomultiplier tubes, activation of discriminators or comparators at the pixel and telescope trigger levels, and digitization of the resultant signals. Each telescope is triggered when the total number of photon-electrons exceeds 30, with a requirement of at least 4 pixels crossing the threshold. Gamma rays are triggered when at least two telescope triggers occur within a short time window, typically spanning several tens of nanoseconds.\par
For each participating telescope in the event reconstruction, image-level selection cuts are applied before Hillas parameter calculation. A standard tail-cut cleaning is first performed to separate Cherenkov shower signals from night-sky background noise. Only cleaned images with a total photo-electron greater than 200 are retained. Additionally, the surviving image must contain at least four non-noise pixels after cleaning, removing faint, fragmented structures dominated by random noise. To avoid systematic bias from image truncation at the camera edge, the centroid of the cleaned shower image is required to lie within 1.5 degrees of the camera centre. \\\indent
\section{Method}
\label{sec3}
High-energy cosmic rays entering the atmosphere generate extensive air showers (EAS) that evolve longitudinally along the shower axis while spreading transversely. Gamma and proton showers show distinct longitudinal and lateral profiles. Gamma rays trigger pure electromagnetic cascades governed by pair production and bremsstrahlung; these showers evolve smoothly, with secondaries tightly clustered near the axis and particle counts fading quickly with atmospheric depth. Consequently, gamma showers feature narrow lateral extents, weak longitudinal tails, and compact Cherenkov images~\cite{hess_crab,gamma-proton-concept}. \par
\begin{figure}[htbp]    
\centering    
\includegraphics[width=1\linewidth]{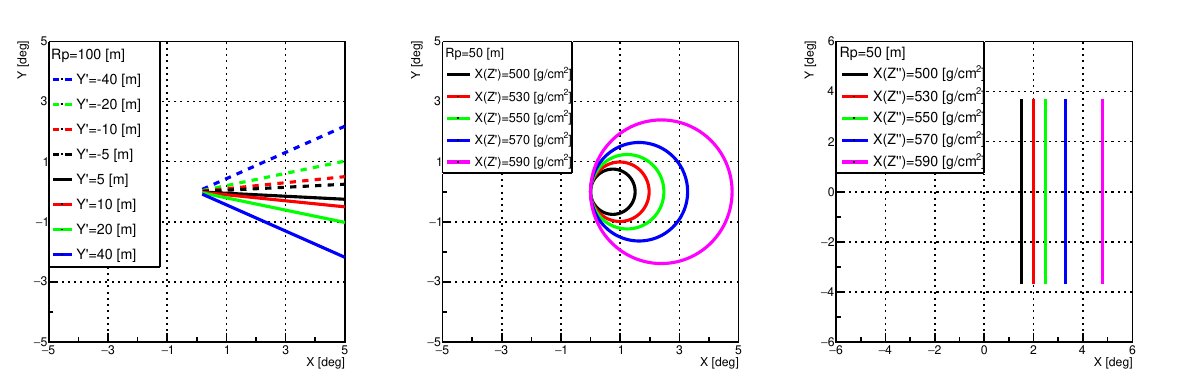}    
\caption{Contour lines of the Y and Z coordinates (converted to atmospheric depth) of Cherenkov photon emission positions projected onto the telescope focal plane. The left panel displays constant-Y contours computed via Eq.~\ref{y-m1}, while the central and right panels show constant-Z contours in the two coordinate systems, calculated from Eq.~\ref{contourz1} and Eq.~\ref{contourz2}, respectively.}    
\label{fig:contour}
\end{figure}

By contrast, proton-driven showers arise from hadronic interactions, generating abundant secondary hadrons, electromagnetic subshowers and muons. This yields complex shower structures, wider lateral extents and large event-wise fluctuations. Muons penetrate deep atmospheric layers with minimal attenuation, creating pronounced tails in longitudinal shower profiles~\cite{muon_IACT}. These disparities in lateral and longitudinal shower evolution form the core physical basis for gamma–proton separation. Spaceborne calorimeters already leverage such morphological differences between electromagnetic and hadronic cascades to perform reliable electron–proton discrimination. For example, the DAMPE experiment uses two core discrimination observables: longitudinal shower leakage (fraction of energy deposited in the final calorimeter layer relative to total shower energy) and transverse shower spread (average shower width measured across calorimeter layers)~\cite{ATIC-ep,DAMPE-ep}. \\\indent
Driven by these motivations, we reconstruct Cherenkov photon emission spatial distributions from IACT shower images and propose observables analogous to those used in DAMPE experiment. We then evaluate their performance for ground-based gamma–proton separation.\\\indent
\begin{figure}[htbp]
    \centering
    \begin{subfigure}[b]{0.48\linewidth}
        \centering
        \includegraphics[width=\linewidth]{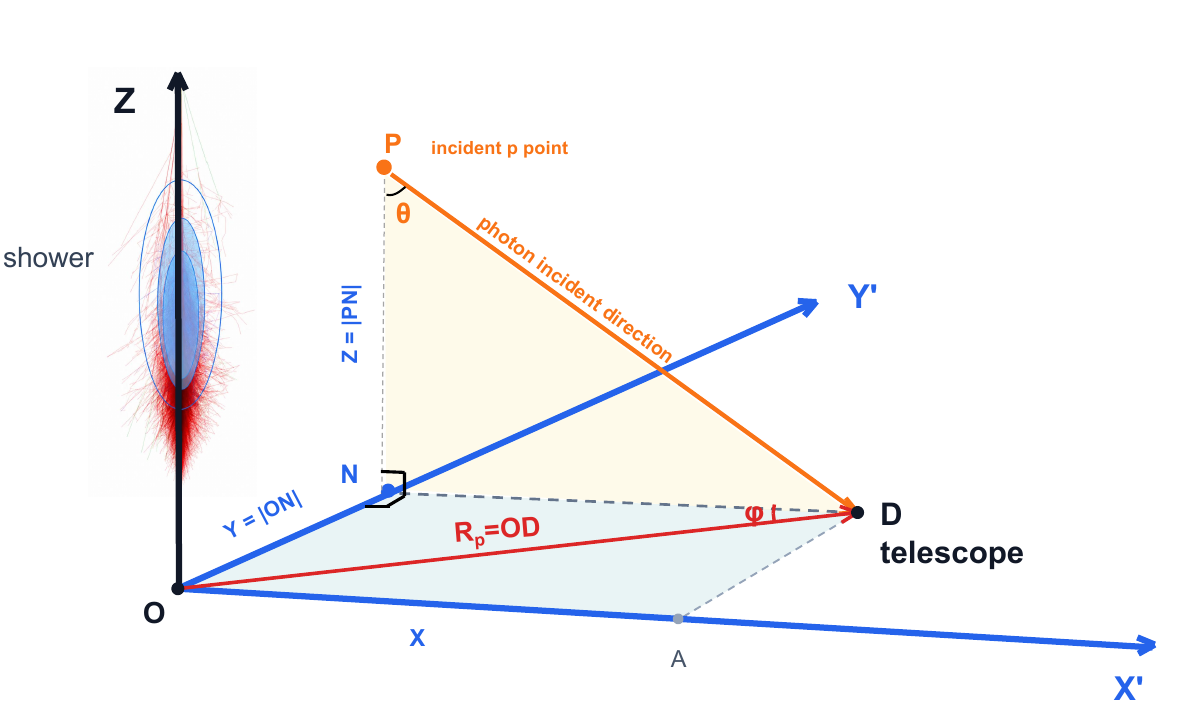}
        \caption{}
        \label{fig:shower1}
    \end{subfigure}
    \hfill
    \begin{subfigure}[b]{0.48\linewidth}
        \centering
        \includegraphics[width=\linewidth]{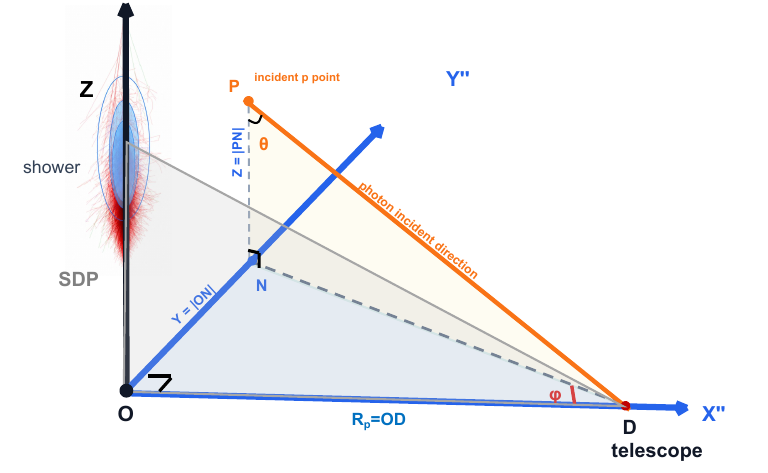}
        \caption{}
        \label{fig:shower2}
    \end{subfigure}
    \caption{Two coordinate systems for reconstructing Cherenkov photon emission positions. The Z axis coincides with the shower axis; D marks the telescope location, and \(\overrightarrow{PD}\) is the trajectory of a Cherenkov photon. The plane PND contains the shower axis and the photon path. Left panel: the Y-axis is perpendicular to plane PND, while the X-axis is orthogonal to both the Z-axis and Y-axis. Right panel: the X-axis points toward the telescope position, and the Y-axis is perpendicular to the Z-axis and X-axis. Cherenkov photon emission positions are reconstructed by projecting photon trajectories onto the Y–Z plane at point P.}
    \label{fig:shower_comparison}
\end{figure}
Cherenkov photon emission positions are reconstructed by tracing Cherenkov photon trajectories onto a reference plane. Two coordinate frames for this plane are shown in Fig.~\ref{fig:shower_comparison}; both use the shower axis as the Z-axis. The difference between Fig.~\ref{fig:shower_comparison}a and Fig.~\ref{fig:shower_comparison}b lies in their X and Y axes: one is defined by photon direction, the other by telescope position. Point D denotes the telescope that detected Cherenkov photons, and the origin O is set so that segment DO (length $R_{p}$) lies perpendicular to the shower axis. The orange arrow shows the photon propagation direction, and the orange shaded plane DPN spans the shower axis and the photon’s path, with PN parallel to the shower axis.\\\indent
In Fig.~\ref{fig:shower1}, The $Y'$-axis (or $\overrightarrow{ON}$) originating at point O is perpendicular to plane DPN and intersects this plane at point N. The Cherenkov photon emission point P corresponds to the intersection of the Cherenkov photon trajectory and the $Y'-Z$ plane. The $X'$-axis (parallel to $\overrightarrow{ND}$) is orthogonal to both the $Y'$-axis and the $Z$-axis. The angle $\varphi$ is the azimuth angle of OD with respect to the $X'$-axis, which is also the angle between DPN and SDP (defined later), the angle $\theta$ is the angle of Cherenkov photon with respect to negative $Z$-axis. Within this coordinate system, the x-coordinate of point P is identically zero. Its y-coordinate equals the length of segment NO, which can be derived via Eq.~\ref{y-m1}; the z-coordinate corresponds to the length of segment PN, obtainable from Eq.~\ref{z-m1}. The Cherenkov photon ultimately reaches the telescope focal plane at point $P'$ (shown in Fig.~\ref{fig:plane image}). The shower detector plane (SDP), bounded by the shower axis and telescope position, intersects the camera focal plane along a straight line. This line forms the Cherenkov image’s principal axis, defined as the new x-axis in Fig.~\ref{fig:plane image} and aligned with the image orientation. The coordinate origin lies at point $O'$ (Fig.~\ref{fig:plane image}), oriented along the reconstructed shower axis with method in reference~\cite{dirrec_st}. In this transformed coordinate system, the polar coordinates of $P'$ on the focal plane are ($\rho$,$\phi$). Where $\rho$ is the angular distance on the focal plane between the photon direction and the shower axis, and $\phi$ is the angle between $O'P'$ and new x-axis — also the angle between SDP and DPN plane. So $\rho$ and $\phi$ are determined by Eq.~\ref{imager} and Eq.~\ref{imagep}, respectively. Combining Eq.~\ref{z-m1}, \ref{imager}, and \ref{imagep} yields Eq.~\ref{contourz1}. Based on Eq.~\ref{contourz1} and Eq.~\ref{y-m1}, the z-coordinate contours of Cherenkov emission positions form focal-plane circles centered at ($R_{p}/(2Z')$,0) with radius $R_{p}/(2Z')$. Meanwhile, emission-position y-coordinate contours represent azimuthal contours across the focal plane, as shown in the left and middle panels of Fig.~\ref{fig:contour}. \\\indent
\begin{figure}[htbp]
    \centering
    \includegraphics[width=1\linewidth]{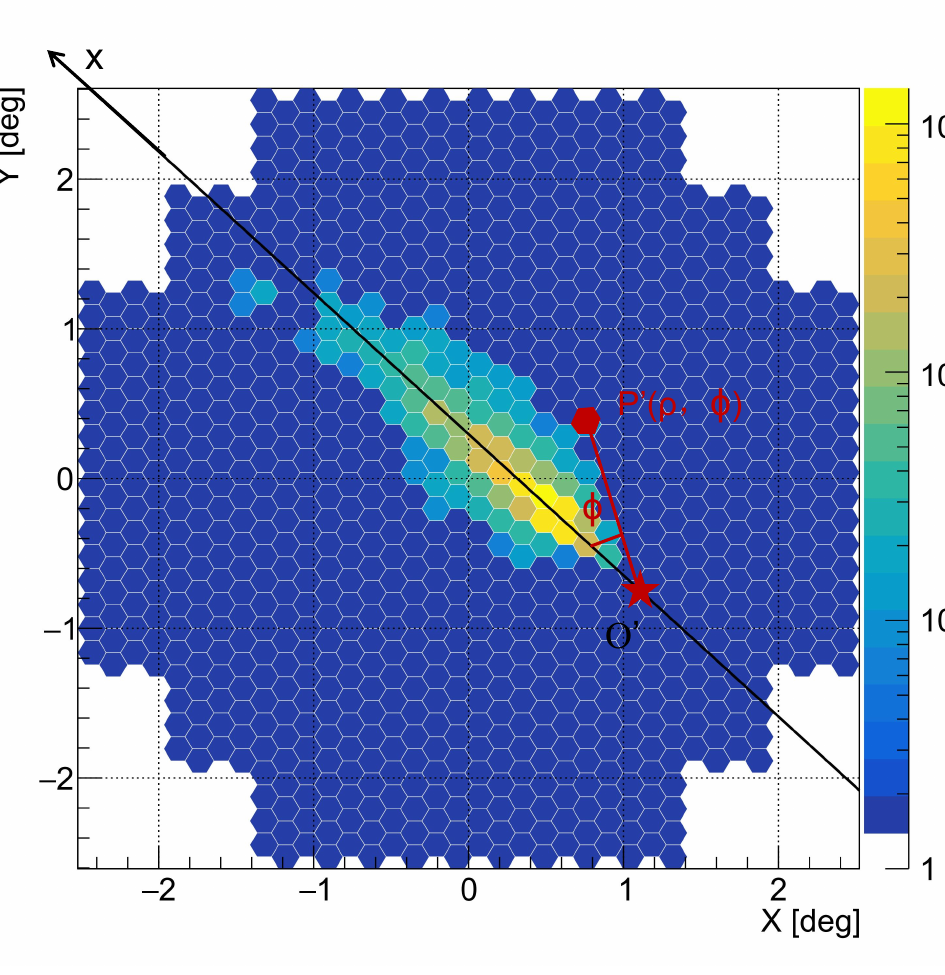}
    \caption{Schematic illustration of Cherenkov-photon imaging on the telescope focal plane and its correlation with the vectors defined in Fig.~\ref{fig:shower_comparison}. \(O'\) denotes the reconstructed shower axis, the black solid line marks the reconstructed shower detector plane, and \(P'\) represents the photon trajectory \(\overrightarrow{PD}\) from Fig.~\ref{fig:shower_comparison}.}
    \label{fig:plane image}
\end{figure}

\begin{equation}
Y'=R_{p}\sin(\varphi) .
\label{y-m1}
\end{equation}
\begin{equation}
Z'=\frac{R_{p}\cos(\varphi)}{\tan(\theta)} .
\label{z-m1}
\end{equation}
\begin{equation}
\rho=\tan\theta
\label{imager}
\end{equation}
\begin{equation}
\phi=\varphi
\label{imagep}
\end{equation}
\begin{equation}
(\rho\cos(\phi)-\frac{R_{p}}{2Z'})^{2}+(\rho\sin(\phi))^2=(\frac{R_{p}}{2Z'})^2
\label{contourz1}
\end{equation}
\\\indent
In Fig.~\ref{fig:shower2}, the $X''$-axis from origin O points toward the telescope($\overrightarrow{OD}$), and the $X''-Z$ plane forms the SDP. The $Y''$-axis through O lies perpendicular to this plane($\overrightarrow{ON}$). The Cherenkov emission point P is reconstructed as the intersection of the photon path and the $Y''-Z$ plane; the DPN plane crosses the $Y''$-axis at point N. $\varphi$ denotes the angle between SDP and DPN plane, which is also the angle between $O'P'$ and new x-axis on the focal plane. $\theta$ is also the angle of Cherenkov photon with respect to negative $Z$-axis. Within this coordinate frame, P has x=0, its y-coordinate equals the length of ON (Eq.~\ref{y-m2}), and its z-coordinate matches the length of PN (Eq.~\ref{z-m2}). Combining Eq.~\ref{z-m2}, \ref{imager}, and \ref{imagep} yields Eq.~\ref{contourz2}. According to Eq.~\ref{contourz2} and~\ref{y-m2}, z-coordinate contours of Cherenkov emission positions form vertical lines at $x=R_{p}/Z''$ on the focal plane, as shown in the right panel of Fig.~\ref{fig:shower1}. Y-coordinate contours are azimuthal contours across the focal plane and closely resemble those in the left panel of Fig.~\ref{fig:shower1}. \\\indent
\begin{equation}
Y''=R_{p}\tan(\varphi) .
\label{y-m2}
\end{equation}
\begin{equation}
Z''=\frac{R_{p}/\cos(\varphi)}{\tan(\theta)} .
\label{z-m2}
\end{equation}
\begin{equation}
\rho cos(\phi)=\frac{R_{p}}{Z''} .
\label{contourz2}
\end{equation}

Fig.~\ref{fig:contour} displays $Y'$, $Z'$, and $Z''$ contours on the focal plane; the $Z'$ and $Z''$ coordinates are converted to atmospheric depth via the U.S. Standard Atmosphere model, which projects the Y–Z shower spatial distribution onto the telescope’s 2D focal plane. For a fixed shower impact distance $R_p = 100~\mathrm{m}$, the left panel of Fig.~\ref{fig:contour} shows mapped contours of the transverse emission y-coordinate, demonstrating that y variations manifest purely as azimuthal shifts across the focal plane. The middle and right panels of Fig.~\ref{fig:contour} show vertical emission-depth contours at different atmospheric depths, corresponding to the two coordinate frames in Fig.~\ref{fig:shower1} and Fig.~\ref{fig:shower2}, respectively. As shown, the contour for 590$\mathrm{g/cm^2}$ forms a circle with angular radius $2.4^\circ$ or a vertical line at $x=4.8^\circ$. \\\indent
Based on the established ground-based coordinate system, the spatial geometry of Cherenkov photons relative to the shower axis can be reconstructed from their Cherenkov image recorded by the telescopes, together with the reconstructed shower direction and core position. Motivated by the discrimination variables employed in the DAMPE experiment, we adopt the average shower spread (denoted $RMS_y$, Eq. \ref{eq:rmsy}) and longitudinal shower leakage (denoted $F_z$, Eq. \ref{eq:Fz}) as new discrimination variables for each triggered telescope. \\\indent
\begin{equation}
RMS_y=\sqrt{\frac{\sum_{i=1}^{n}(y_i-\bar{y})^2N_i^{pe}}{\sum_{i=1}^{n}N_i^{pe}}}
\label{eq:rmsy}
\end{equation}
\begin{equation}
\bar{y}=\frac{\sum_{i=1}^{n}y_iN_i^{pe}}{\sum_{i=1}^{n}N_i^{pe}}
\label{eq:meany}
\end{equation}
\\\indent
In Eq.\ref{eq:rmsy}, i denotes the index of the triggered pixel, n the total number of triggered pixels, $y_i$ the y-coordinate ($Y'$ or $Y''$) of the $i^\text{th}$ triggered pixel calculated via Eq.\ref{y-m1} or Eq.\ref{y-m2}, and $N_i^\text{pe}$ the number of photoelectrons detected by the $i^\text{th}$ triggered pixel, with $\bar{y}$ the weighted mean of the y-coordinate (Eq.\ref{eq:meany}). $RMS_y$ thus characterizes differences in the transverse spatial development of air showers and acts as an observable for gamma–proton discrimination.\\\indent
\begin{equation}
F_z=\frac{N_{\mathrm{zcut}}^{pe}}{N_{\mathrm{all}}^{pe}}
\label{eq:Fz}
\end{equation}
\\\indent
In Eq. \ref{eq:Fz}, $N_{\mathrm{all}}^{pe}$ denotes the total Cherenkov photoelectrons per telescope, while $N_{\mathrm{zcut}}^{pe}$ counts photoelectrons confined within selected z-coordinate contour regions ($Z'$ or $Z''$ above reference values), as illustrated in Fig. \ref{fig:contour}. $F_z$ thus captures variations in the longitudinal air-shower tail development and serves as an observable for gamma--proton discrimination. \\\indent
We investigated the differences in the $RMS_y$ variable between $Y'$ and $Y''$, as well as the $F_z$ observable computed using $Z'$ and $Z''$. We find that $RMS_y$ from $Y'$ exhibits comparable discrimination performance to $RMS_y$ from $Y''$ for $R_p>$$\sim100$ m. This arises because $sin(\varphi)$ and $tan(\varphi)$ are similar at small $\varphi$ when $R_p$ has large values (see Eq.~\ref{y-m1} and Eq.~\ref{y-m2}), whereas $RMS_y$ from $Y'$ yields superior gamma–proton discrimination performance relative to $Y''$ for $R_p<$$\sim$100 m. Preliminary tests show weak gamma--proton separation power for $F_z$ derived from both $Z'$ and $Z''$, which may arise from the limited field of view of the H.E.S.S. telescopes. We thus restrict all subsequent analysis to $RMS_y$ calculated with $Y'$ and evaluate its potential to improve gamma--proton discrimination.



To evaluate the discrimination performance of $RMS_y$, we define the gamma--proton discrimination metric ($DA$, hereafter) via Eq.~\ref{eq:DA}, with the gamma-ray selection efficiency fixed at approximately $90\%$. Here, $\epsilon_\gamma$ and $\epsilon_{\rm CR}$ denote the gamma-ray selection efficiency and the surviving fraction of cosmic-ray protons, respectively. A larger value of $DA$ corresponds to a lower proton survival efficiency and therefore a better gamma--proton discrimination performance.
\begin{equation}
DA = -\log_{10}\left(\epsilon_{\rm CR}\right)
\bigg|_{\epsilon_\gamma \simeq 0.90}
\label{eq:DA}
\end{equation}

\section{Result}
\label{sec4}
\subsection{ Single telescope}
\label{subsec3}
\begin{figure}[H]
    \centering
    \includegraphics[width=1\linewidth]{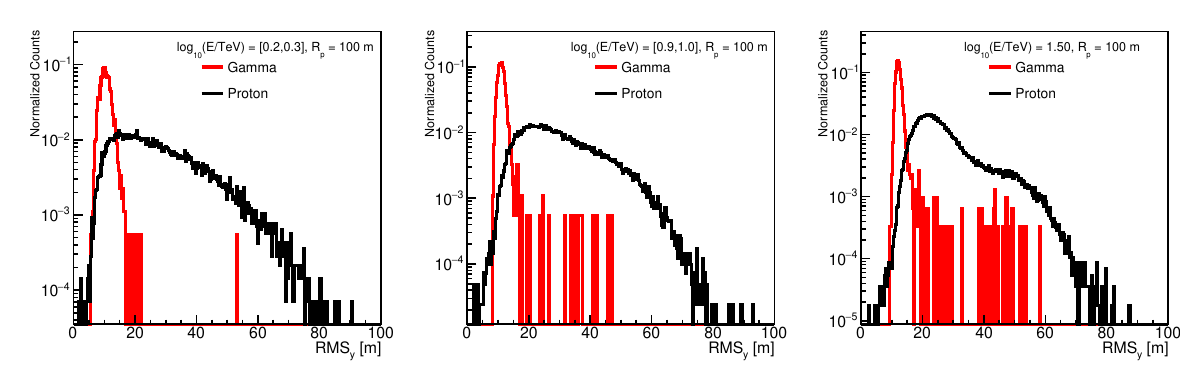}
    \caption{One-dimensional $RMS_y$ distributions of $\gamma$-ray (red) and proton-induced (black) events at $R_p = 100~\mathrm{m}$ for $\log_{10}(E/\mathrm{TeV}) = [0.2,0.3]$ (left), [0.9,1.0] (middle), and 1.5 (right).}
    \label{fig:RMSy_distribution}
\end{figure}

\begin{figure}[H]
    \centering
    \includegraphics[width=1\linewidth]{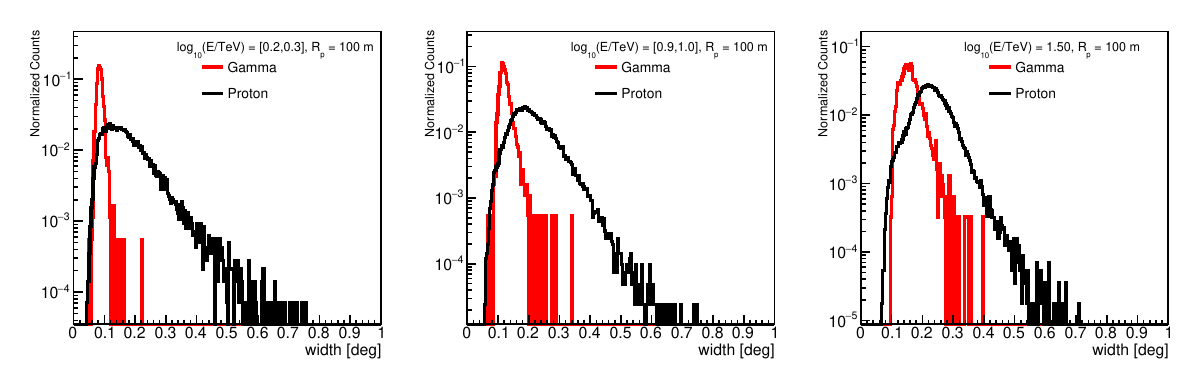}
    \caption{Comparison of the one-dimensional \textit{width} distributions of $\gamma$-ray (red) and proton-induced (black) events at $R_p = 100~\mathrm{m}$. Panels correspond to $\log_{10}(E/\mathrm{TeV}) = [0.2,0.3]$ (left), [0.9,1.0] (middle), and 1.5 (right).}
    \label{fig:Width_distribution}
\end{figure}

The Hillas parameter \textit{width} describes the lateral extent of Cherenkov images and serves as a standard discriminator for traditional IACT gamma--proton separation. To compare $RMS_y$ and \textit{width}, we examine one-dimensional gamma-ray and cosmic-ray proton event distributions at fixed $R_p$ across three energy bins: $\log_{10}(E/\mathrm{TeV})=[0.2,0.3]$, $[0.9,1.0]$, and $1.5$. These distributions are displayed in Figs.~\ref{fig:RMSy_distribution} and \ref{fig:Width_distribution}. As shown in Fig.~\ref{fig:RMSy_distribution}, the $RMS_y$ distribution is narrowly concentrated around 10 meter for gamma rays, while cosmic-ray protons exhibit a broader distribution with a pronounced right-tail effect.\par

\begin{figure}[H]
    \centering
    \includegraphics[width=1\linewidth]{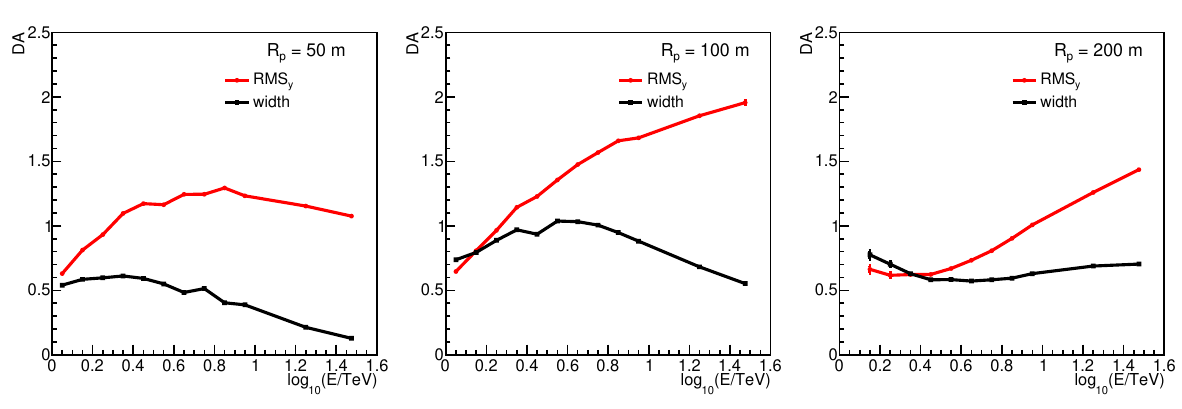}
    \caption{Energy dependence of the discrimination ability (DA) for $RMS_y$ and \textit{width} at fixed $R_p$. The left, middle, and right panels correspond to $r_p = 50~\mathrm{m}$, $100~\mathrm{m}$, and $200~\mathrm{m}$, respectively.}
    \label{fig:DA_single_E}
\end{figure}

To quantitatively compare the discrimination power of $RMS_y$ and \textit{width}, we examine the energy dependence of discrimination metric DA at fixed $R_p$ (Fig.~\ref{fig:DA_single_E}). The DA of $RMS_y$ rises with energy, and surpasses width substantially at small $R_p$ and high energies. This proves $RMS_y$ better captures transverse shower variations at high energies, yielding superior gamma–proton separation performance.\par
\begin{figure}[H]
    \centering
    \includegraphics[width=1\linewidth]{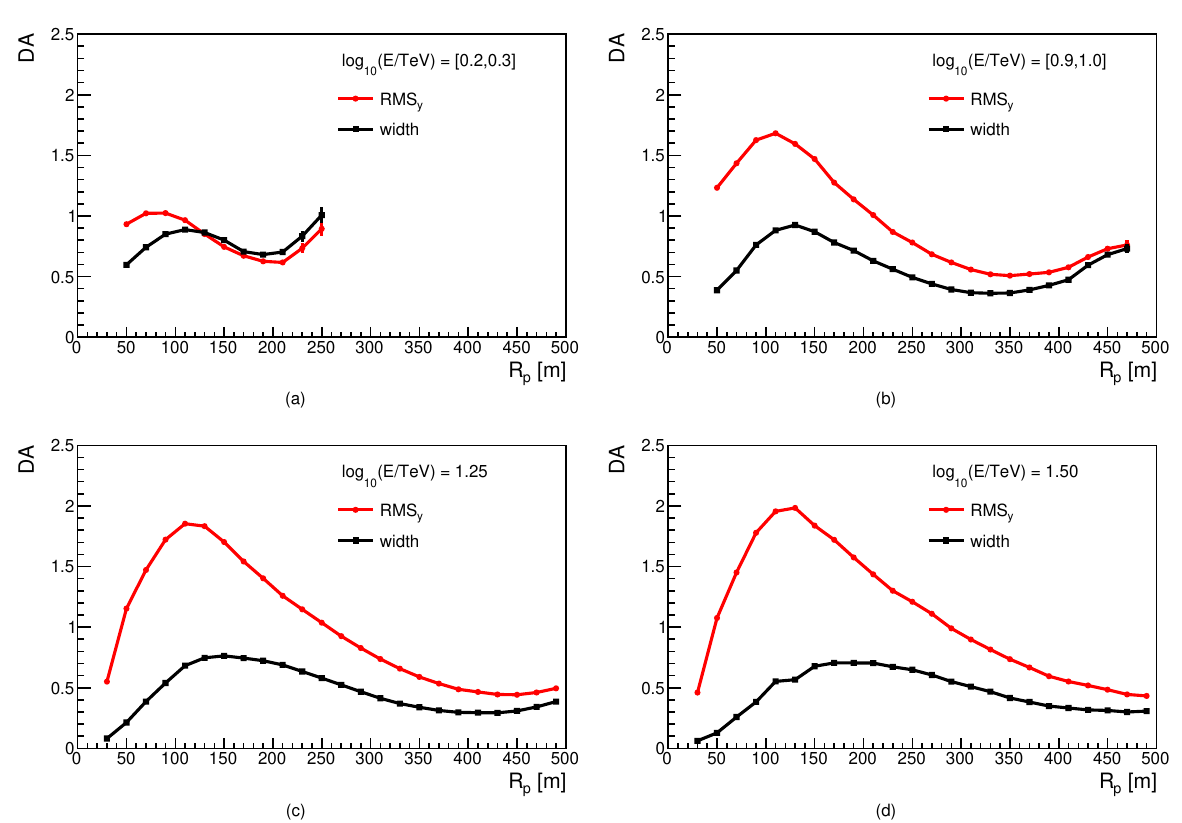}
    \caption{Discrimination ability (DA) of $RMS_y$ and \textit{width} as a function of shower impact distance $R_p$ at four representative energies: $\log_{10}(E/\mathrm{TeV}) = [0.2,0.3]$ (a), [0.9,1.0] (b), 1.25 (c), and 1.5 (d).}
    \label{fig:DA_single_Rp}
\end{figure}

To further study how the discrimination performance varies with $R_p$ for the two observables, we select four representative energy bins: $\log_{10}(E/\mathrm{TeV})=[0.2,0.3]$, $[0.9,1.0]$, 1.25, and 1.5, and systematically compare the discrimination metric $DA$ of $RMS_y$ and \textit{width} as functions of $R_p$. As shown in Fig.~\ref{fig:DA_single_Rp}, both observables follow similar trends with $R_p$: $DA$ gradually increases with $R_p$, reaches a maximum at approximately $R_p = 100$--$130,\mathrm{m}$, and then slowly declines. This indicates that the discrimination performance of single-telescope observables depends strongly on the geometric position of the telescope relative to the shower axis, namely the shower impact distance.
At low energies $log_{10}(E/\mathrm{TeV})<$0.3, $RMS_y$ yields discrimination performance comparable to \textit{width}, yet it outperforms \textit{width} at higher energies. Above $10\,\mathrm{TeV}$, $RMS_y$ provides markedly superior gamma–proton separation relative to the conventional \textit{width} observable.\par
As discussed in Section \ref{sec_disc}, \(RMS_y\) outperforms \textit{width} in discrimination for single telescope mainly because \textit{width} strongly depends on each telescope’s Cherenkov image \textit{size} (total recorded Cherenkov photon-electron in one telescope’s image), which broadens the \textit{width} distribution without \textit{size}-dependent corrections, whereas \(RMS_y\) is less sensitive to Cherenkov image \textit{size}. \\\indent

\subsection{Multiple telescopes}
\label{subsec4}
Multi-telescope IACT arrays record a single air shower from distinct viewing angles and impact distances $R_p$ requiring event-wise fusion of discrimination information from each telescope. Standard Hillas analysis employs the array observable mean reduced scaled width(or MRSW)~\cite{hess_crab}: it normalizes \textit{width} to uniform mean and RMS across all $R_p$ and \textit{size} bins before averaging over triggered telescopes. According to the Neyman–Pearson-Lemma the best test to decide between two hypotheses is the likelihood ratio~\cite{LLR}. While widely adopted for gamma-ray source detection~\cite{tools_iact}, it is rarely applied to enhance gamma–proton separation in ground-based IACTs, unlike its established use in space-borne experiments~\cite{TRD_LLR}. We further adopt the likelihood-ratio method to combine single-telescope observables.Therefore, we further combine the single-telescope discrimination observables using both the likelihood-ratio method and the mean reduced scaled-based combination approach.\par

\begin{equation}
\mathrm{LR}_{i}(R_p,size)=\frac{P_{\gamma}(x_i \mid R_p,size)}{P_{p}(x_i \mid R_p,size)} .
\label{eq:single_tel_lr}
\end{equation}
\begin{equation}
LLR=\sum_{i=1}^{N_{\mathrm{tel}}}
\ln\left(\mathrm{LR}_{i}(Rp,size)\right) .
\label{eq:array_lr}
\end{equation}

For each event, the likelihood ratio LR of the $i^\text{th}$ telescope is defined by Eq.~\ref{eq:single_tel_lr}. Here, $P_{\gamma}(x_i \mid R_p,\mathit{size})$ gives the probability of the $i^\text{th}$ telescope observable $x_i$ under the gamma hypothesis at the telescope’s measured $R_p$ and \textit{size}, and $P_{p}(x_i \mid R_p,\mathit{size})$ corresponds to the cosmic-ray proton hypothesis. Probability densities per $R_p$ and \textit{size} bin are obtained from simulated samples, and the combined log-likelihood ratio LLR is defined via Eq.~\ref{eq:array_lr}, with $N_{\mathrm{tel}}$ representing the count of triggered telescopes.\par
We define three combined observables from different single-telescope quantities and merging strategies: MRSW, $\mathrm{LnY}$, and $\mathrm{LnW}$. Here, $\mathrm{LnY}$ denotes the LLR of $RMS_y$, while $\mathrm{LnW}$ corresponds to the LLR of \textit{width}. We further evaluate the discrimination performance of these three observables for mutual comparison. For this comparison, we select a circular core region centered at the H.E.S.S. array, with the circle radius (denoted as $\mathrm{R_c}$) as the free parameter.\par
\begin{figure}[H]
    \centering
    \includegraphics[width=\linewidth]{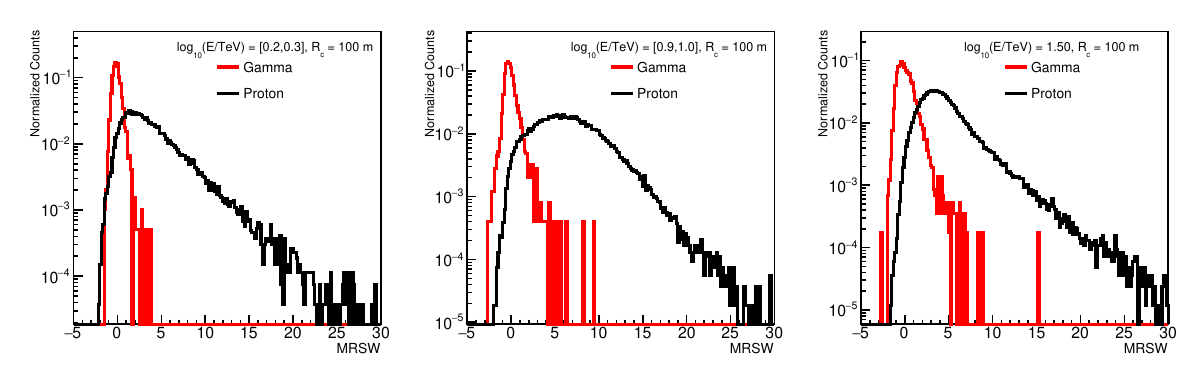}
    \caption{Comparison of the one-dimensional MRSW distributions for $\gamma$-ray (red) and proton-induced (black) events with $R_c = 100~\mathrm{m}$. The left, middle, and right panels correspond to $\log_{10}(E/\mathrm{TeV}) = [0.2,0.3]$, $[0.9,1.0]$, and $1.5$, respectively.}
    \label{fig:MRSW}
\end{figure}

\begin{figure}[H]
    \centering
    \includegraphics[width=\linewidth]{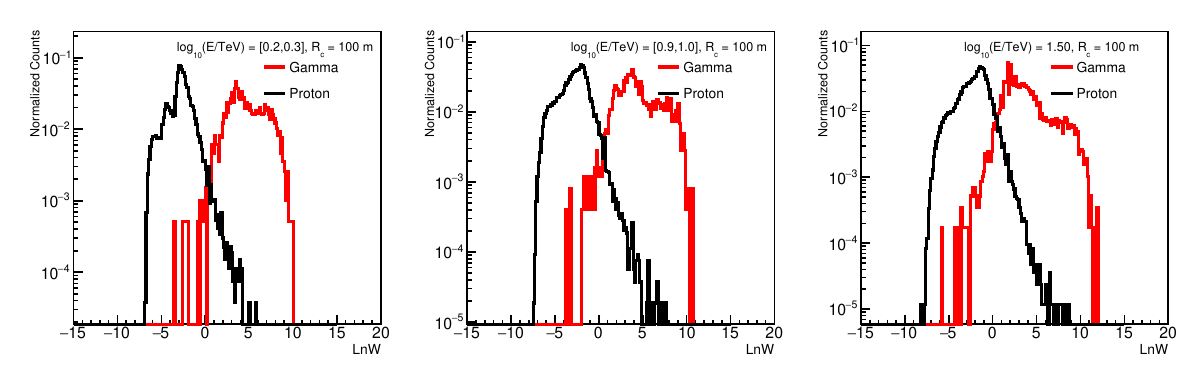}
    \caption{Comparison of the one-dimensional $\mathrm{LnW}$ distributions for $\gamma$-ray (red) and proton-induced (black) events with $R_c = 100~\mathrm{m}$. The left, middle, and right panels correspond to $\log_{10}(E/\mathrm{TeV}) = [0.2,0.3]$, $[0.9,1.0]$, and $1.5$, respectively.}
    \label{fig:LnW}
\end{figure}

\begin{figure}[H]
    \centering
    \includegraphics[width=\linewidth]{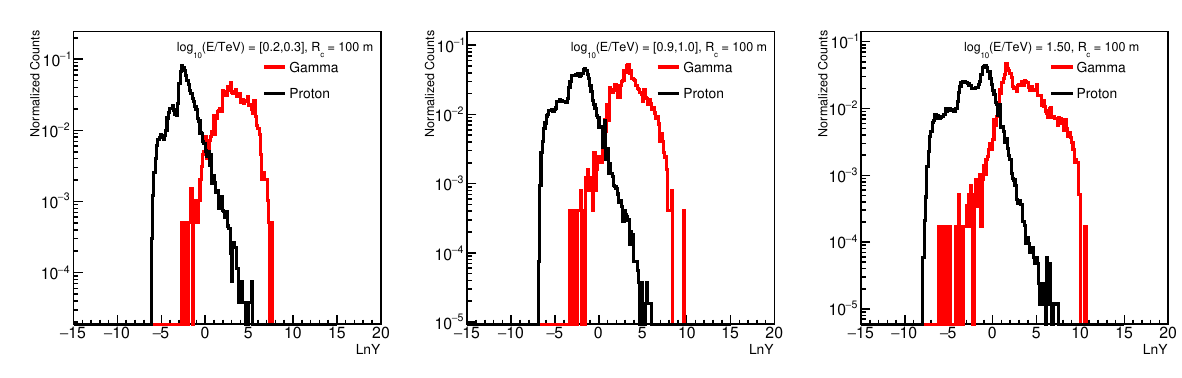}
    \caption{Comparison of the one-dimensional $\mathrm{LnY}$ distributions for $\gamma$-ray (red) and proton-induced (black) events with $R_c = 100~\mathrm{m}$. The left, middle, and right panels correspond to $\log_{10}(E/\mathrm{TeV}) = [0.2,0.3]$, $[0.9,1.0]$, and $1.5$, respectively.}
    \label{fig:LnY}
\end{figure}
\begin{figure}[H]
    \centering
    \includegraphics[width=1\linewidth]{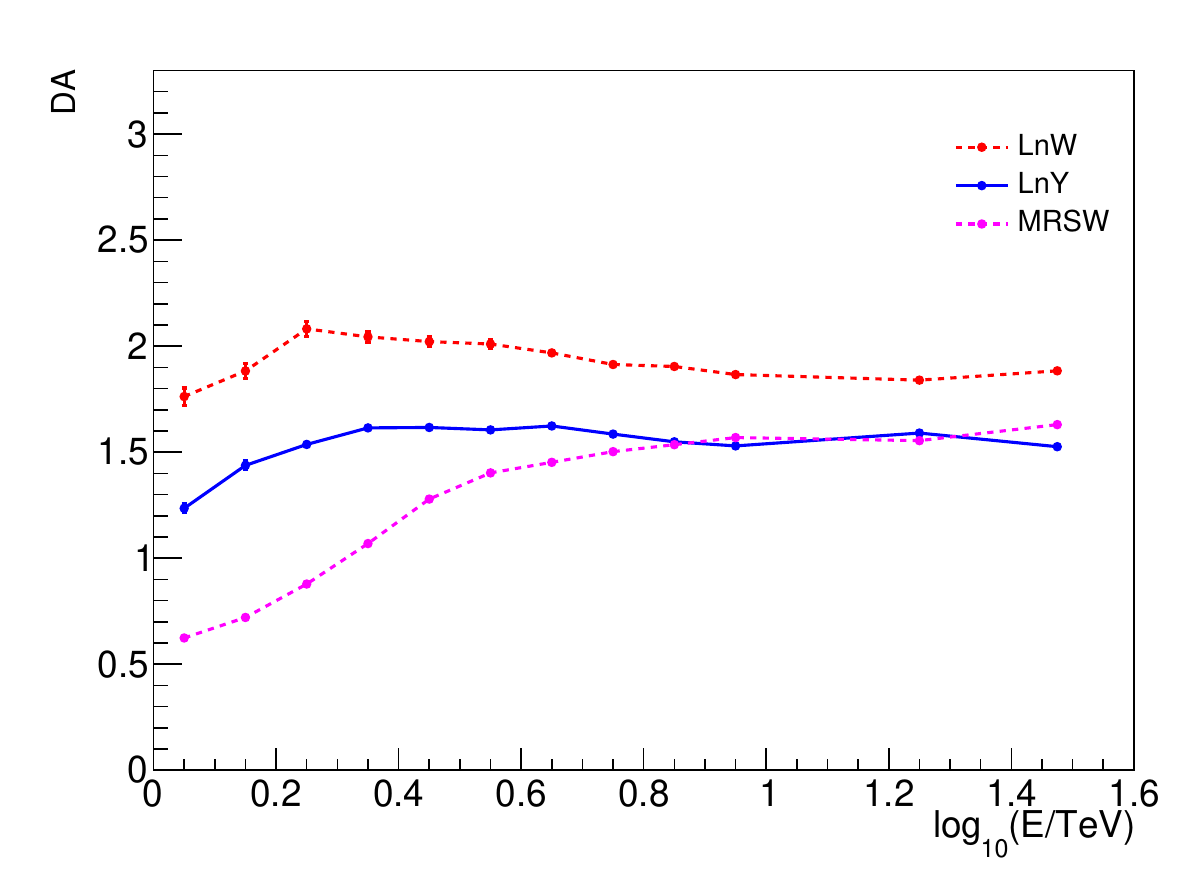}
    \caption{Comparison of the energy dependence of discrimination ability (DA) for \(\mathrm{MRSW}\) (magenta dashed line), \(\mathrm{LnW}\) (red dashed line), and \(\mathrm{LnY}\) (blue solid line). }
    \label{fig:DA_comb_Rc500}
\end{figure}

Figure~\ref{fig:MRSW},  \ref{fig:LnW}, and \ref{fig:LnY} show the one-dimensional distributions of MRSW, $\mathrm{LnW}$, and $\mathrm{LnY}$ at fixed $R_c=100\ \mathrm{m}$ for $\log_{10}(E/\mathrm{TeV})=[0.2,0.3]$, $[0.9,1.0]$, and 1.5, respectively. Figure~\ref{fig:DA_comb_Rc500} presents the energy evolution of the discrimination performance $DA$ for the three array-level observables using all triggered events.\par
As observed, $\mathrm{LnW}$ substantially outperforms MRSW and $\mathrm{LnY}$, particularly at low primary energies. Specifically, near $1~\mathrm{TeV}$, $\mathrm{LnW}$ yields an improvement of about one order of magnitude relative to the conventionally used variable $\mathrm{MRSW}$; above $10~\mathrm{TeV}$, this enhancement drops to more than 50\% compared with $\mathrm{MRSW}$. In addition, \(\mathrm{LnY}\) also outperforms \(\mathrm{MRSW}\) at low energies (by roughly a factor of 4 near 1 TeV), while its performance is comparable to \(\mathrm{MRSW}\) above $\sim$6 TeV. These results demonstrate that $\mathrm{LnW}$ provides superior gamma--proton discrimination over the investigated energy range. Furthermore, the likelihood-ratio combination outperforms the conventional mean reduced scaled scheme, demonstrating its greater ability to merge discrimination information from multiple telescopes. \\\indent
It is also important to note that the MC sample used to construct the probability distributions of \textit{width} and \(RMS_y\) differs from the sample used to produce Fig.~\ref{fig:DA_comb_Rc500}, ensuring they are uncorrelated. Furthermore, the likelihood-ratio method is straightforward to implement: it does not require uniform performance across all telescopes, a condition often unmet for complex arrays composed of multiple telescope types. \par

\begin{figure}[H]
    \centering
    \includegraphics[width=1\linewidth]{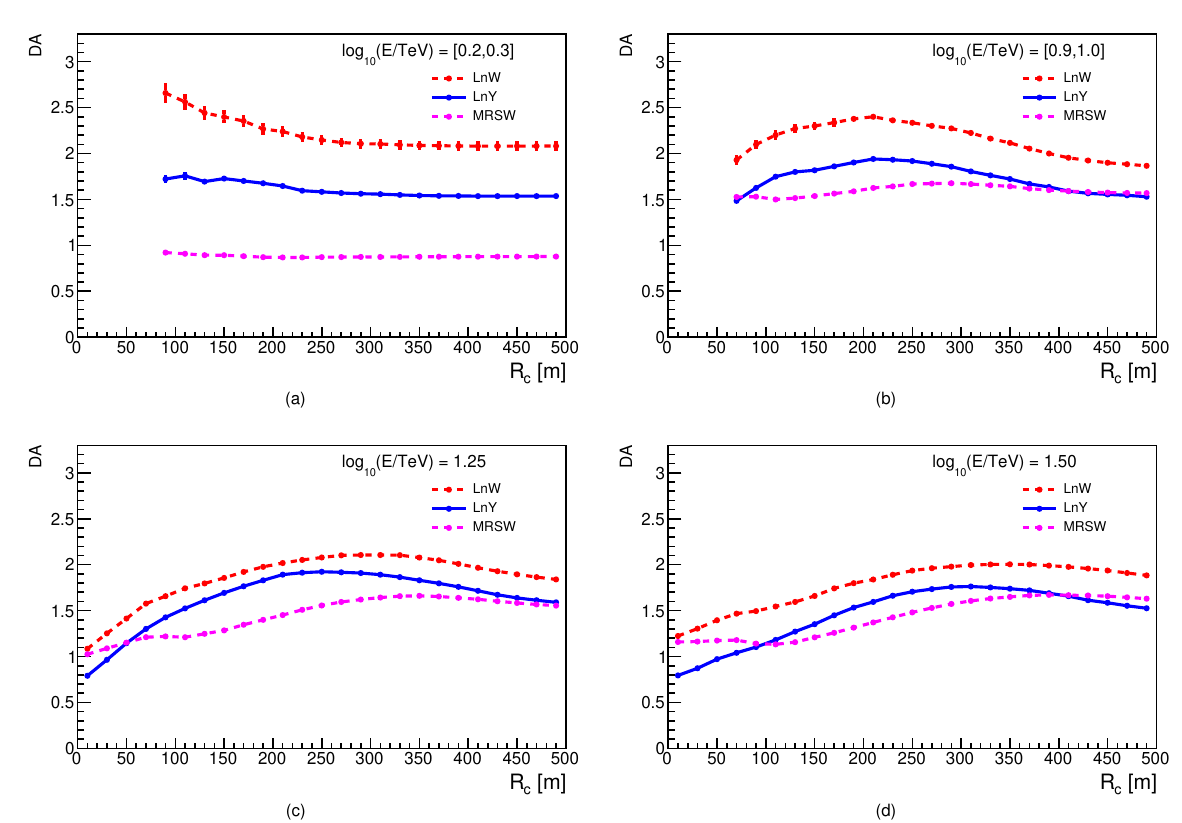}
    \caption{Discrimination ability (DA) of MRSW, $\mathrm{LnY}$, and $\mathrm{LnW}$ as a function of $\mathrm{R_c}$ in four energy intervals: $\log_{10}(E/\mathrm{TeV})=[0.2,0.3]$(a), $[0.9,1.0$](b, 1.25 (c), and 1.5(d). }
    \label{fig:DA_comb_Rc}
\end{figure}

We further study how the discrimination performance of the three combined observables varies with the circle radius $R_c$ at fixed energy bins in Fig.~\ref{fig:DA_comb_Rc}. The four representative energy bins are $\log_{10}(E/\mathrm{TeV})=[0.2,0.3]$, $[0.9,1.0]$, 1.25, and 1.5. The findings are consistent with those obtained from Fig.~\ref{fig:DA_comb_Rc500}. The lower bound of \(R_c\) for the lowest-energy bin begins at 70 m to accumulate sufficient events, compensating for the low trigger efficiency.\par 


\section{Discussion}
\label{sec_disc}
\begin{figure}[H]
    \centering
    \includegraphics[width=1\linewidth]{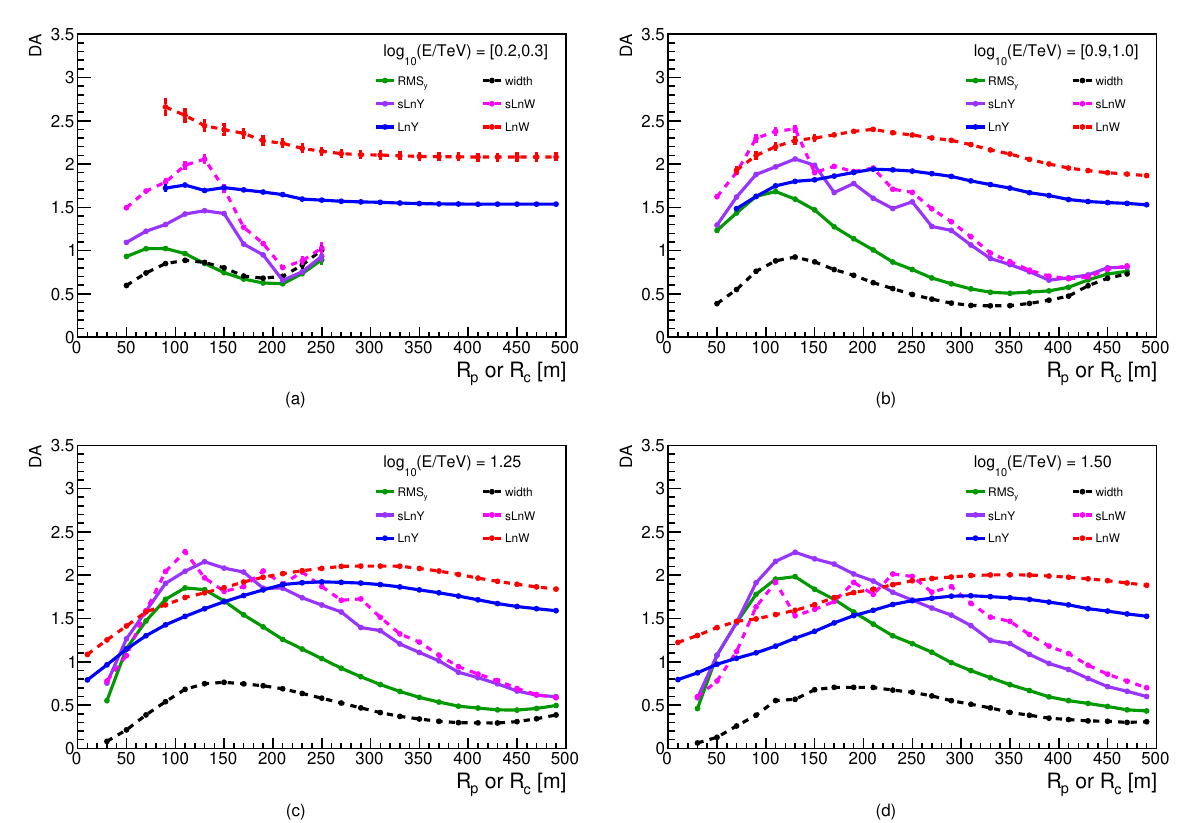}
    \caption{Discrimination ability (DA) of $RMS_y$, $\mathrm{width}$, single-telescope likelihood ratios $\mathrm{sLnY}$ and $\mathrm{sLnW}$, and multi-telescope combined likelihood ratios $\mathrm{LnY}$ and $\mathrm{LnW}$, as a function of $R_c$ for four representative energy intervals: $\log_{10}(E/\mathrm{TeV})=[0.2,0.3]$ (a), $[0.9,1.0]$ (b), $1.25$ (c), and $1.5$ (d).}
    \label{fig:DA_RpRc}
\end{figure}
From Fig.~\ref{fig:DA_single_Rp} and Fig.~\ref{fig:DA_comb_Rc}, \(RMS_y\) yields better single-telescope discrimination than \textit{width}, while the likelihood-ratio combinations of these observables (\(\mathrm{LnY}\), \(\mathrm{LnW}\)) show the reverse trend. To understand this, we construct single-telescope likelihood-ratio observables $\mathrm{sLnY}$ and $\mathrm{sLnW}$ from \(RMS_y\) and \textit{width}. Their results, plotted as magenta and purple lines in Fig.~\ref{fig:DA_RpRc}, reproduce the behaviour seen in Fig.~\ref{fig:DA_single_Rp} and Fig.~\ref{fig:DA_comb_Rc}. $\mathrm{sLnW}$ greatly improves discrimination over \textit{width}, with only moderate differences between $\mathrm{sLnY}$ and \(RMS_y\). Thus, $\mathrm{sLnW}$ performs slightly better than $\mathrm{sLnY}$, in agreement with the multi-telescope combination result. \\\indent
\begin{figure}[H]
    \centering
    \includegraphics[width=1\linewidth]{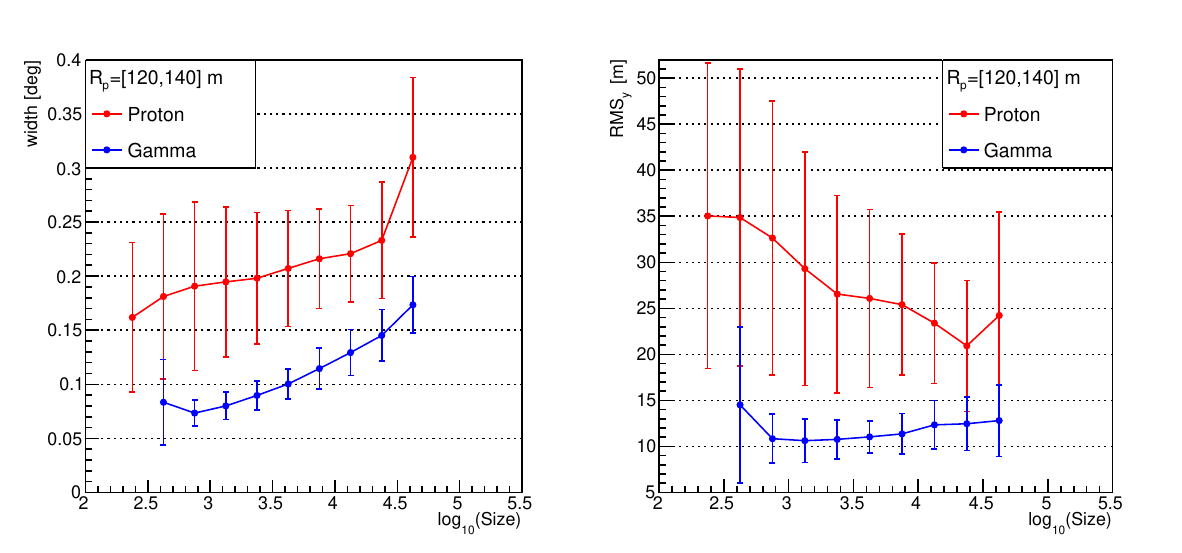}
    \caption{Dependence of \textit{width} (left panel) and \(RMS_y\) (right panel) on Cherenkov image \textit{size} for \(R_p\) ranging from 120 m to 140 m. The vertical coordinate of each point gives the sample mean, with error bars indicating the sample RMS. Red points correspond to proton events, and blue points to gamma-ray events.}
    \label{fig:var_size}
\end{figure}
To further investigate this effect, Fig.~\ref{fig:var_size} presents the variation of \(RMS_y\) and \textit{width} with Cherenkov image \textit{size} for a single \(R_p\) bin. The y-value of each point denotes the sample mean, and error bars correspond to the sample RMS. As shown, \textit{width} exhibits a stronger dependence on \textit{size}, which broadens its distribution and degrades discrimination. In contrast, the likelihood-ratio method and mean reduced scaled \textit{width} perform corrections per \(R_p\) and \textit{size} bin, substantially improving discrimination performance. \\\indent
\section{Conclusion}
\label{sec5}
This work addresses gamma--proton separation for ground-based imaging atmospheric Cherenkov telescope (IACT) experiments. The spatial distribution of Cherenkov photon emission points is reconstructed from recorded shower images. Inspired by discriminator design strategies used in space-borne calorimeters, where the transverse and longitudinal shower profiles are widely exploited for gamma--proton separation, we construct a new ground-based discrimination observable, $RMS_y$, from the transverse spatial distribution of Cherenkov photons in telescope-measured Cherenkov images. We then carry out a comprehensive analysis of the discrimination performance of $RMS_y$.\par
For individual telescopes, $RMS_y$ delivers discrimination performance comparable to the conventional Hillas-based \textit{width} variable at low energies near $\log_{10}(E/\mathrm{TeV})\leq0.3$, and gradually outperforms \textit{width} with increasing energy. Above $10\ \mathrm{TeV}$, $RMS_y$ achieves substantially better gamma--proton separation than the \textit{width} observable, with improvements ranging from tens of percent to over one order of magnitude. This is mainly because \(RMS_y\) is less sensitive to Cherenkov image \textit{size} than \textit{width}.\par
For array-level analysis, we apply the likelihood-ratio method to combine the single-telescope $RMS_y$ and \textit{width} information from multiple telescopes, yielding two new array-level observables: $\mathrm{LnY}$ and $\mathrm{LnW}$. The results show that, around $1~\mathrm{TeV}$, $\mathrm{LnW}$ substantially outperforms $\mathrm{MRSW}$ and $\mathrm{LnY}$ across all $R_c$ ranges, with an improvement of about one order of magnitude compared with $\mathrm{MRSW}$. Above $10~\mathrm{TeV}$, this enhancement decreases to more than 50\% compared with $\mathrm{MRSW}$. \(\mathrm{LnY}\) also outperforms \(\mathrm{MRSW}\) across all \(R_c\) ranges at low energies, improving by roughly a factor of 4 near \(1~\mathrm{TeV}\), and its performance becomes closer to \(\mathrm{MRSW}\) above ~6 TeV. Its performance, however, remains inferior to that of \(\mathrm{LnW}\). \par
In summary, this work demonstrates that discriminator design strategies developed for space-borne calorimeters can be adapted to ground-based IACT experiments, and the newly introduced observable is less sensitive to Cherenkov image \textit{size}. The likelihood-ratio combination outperforms the conventional mean reduced scaled scheme, showing its superior capability to combine discriminant information from multiple telescopes. These approaches improve gamma–proton discrimination for ground-based imaging Cherenkov telescope arrays. \\\indent

\section*{Declarations}
\begin{itemize}
\item Funding
\\\indent
This work was supported in China by the National Natural Science Foundation of China (NSFC, No.12205244).
\item Conflict of interest
\\\indent
On behalf of all authors, the corresponding author states that there is no conflict of interest.
\end{itemize}


\backmatter

\bigskip

\begin{appendices}




\end{appendices}


\bibliography{sn-bibliography}

\end{document}